\documentclass[sn-basic]{sn-jnl}

\usepackage{amsmath,amssymb,amsfonts}
\usepackage{booktabs}
\usepackage{xcolor}
\usepackage{listings}

\usepackage{graphicx}

\usepackage{pgfplots}
\pgfplotsset{compat=1.18}
\pgfplotsset{
  emsefig/.style={
    width=\linewidth, height=5.4cm,
    font=\sffamily\footnotesize,
    symbolic x coords={O0,O1,O2,O3,Os},
    xtick=data,
    xlabel={Compiler optimization level},
    grid=major, grid style={gray!20},
    tick align=outside,
    legend cell align=left,
    legend style={font=\sffamily\scriptsize, draw=gray!50, fill=white,
                  fill opacity=0.9, text opacity=1},
  },
  refined/.style={black, solid, thick, mark=*, mark size=1.9pt},
  rawghidra/.style={gray!65!black, dashed, thick, mark=square*, mark size=1.9pt},
  nullline/.style={black!55, densely dotted, thin, no marks},
}

\newcommand{\Os}{\texttt{-Os}}
\newcommand{\Ozero}{\texttt{-O0}}
\newcommand{\Othree}{\texttt{-O3}}

\begin{document}

\title[A Memorization Floor]{A Memorization Floor for LLM Refinement of Decompiled Code}

\author[1]{\fnm{Muhammad} \sur{Asjad}}\email{m.asjad2401@gmail.com}

\affil[1]{\orgdiv{School of Electrical Engineering and Computer Science
(SEECS)}, \orgname{National University of Sciences and Technology (NUST)},
\orgaddress{\city{Islamabad}, \country{Pakistan}}}

\abstract{We introduce a \emph{memorization floor}: a within-item control separating what
LLM refinement of decompiler output recovers from its input from what it
recovers from its prior. Refine a function, then refine it again from an input
whose identifiers have been destroyed, and measure what survives. Because the
comparison is within-item, corpus difficulty cannot contribute; it costs twenty
API calls. Applied to functions written after our analysis plan was committed,
so no released model could have memorized them, it reports two things.
Recovery is real: refined output sits $+0.072$ to $+0.137$ above an
arm-matched permutation null built from its own output vocabulary. But it does
not depend on the input we ablate: destroying the input's dataflow changes the
naming gain by $+0.001$ (95\% CI $[-0.026, +0.026]$), and removing type
prefixes or permuting names changes it by no more. A second refiner from
another vendor, registered in advance and given byte-identical inputs,
reproduces this --- twelve contrasts, two models, twelve nulls. Readability
stays at ceiling throughout, so a reader is given no signal. The null is
bounded, not absolute: contributions under $0.056$ are invisible, and the
ablation leaves operations intact, so naming from those alone remains a
competing reading. No registered hypothesis was confirmed, and we report the
five instrument failures behind that in full, including a reassembly harness
biased against the treated arm and an equivalence checker we registered without
checking it worked on our inputs.}


\maketitle
\section{Introduction}
\label{sec:intro}

Decompiled C is hard to read. A stripped binary run through Ghidra yields
functions named \texttt{FUN\_00101169}, parameters named \texttt{param\_1},
locals named \texttt{iVar1}, and types named \texttt{undefined4}. A large
language model asked to refine that output produces something that looks like
source code: descriptive names, explanatory comments, plausible types. The
transformation is dramatic, and close to saturated --- a blind judge rates raw
Ghidra output $1.46$ out of $5$ for usefulness and the refined output $4.89$.

This is what makes refinement dangerous. Refined output carries no marker
distinguishing a name recovered by analyzing dataflow from a name recalled
because the function's shape is familiar. Both arrive as confident,
well-commented C. A reverse engineer has no way to tell which they are reading,
and the readability gain actively encourages trust. Readable-but-wrong is the
most dangerous output a reverse engineer can receive, and readability metrics
by construction cannot detect it.

The concern is concrete. Decompilation benchmarks derived from HumanEval are
documented as contaminated in code-LLM training
data~\citep{riddell2024contamination}, and the functions used to evaluate
refinement are overwhelmingly canonical: Fibonacci, binary search, matrix
multiply. In an earlier pilot of this pipeline we watched the refinement pass
recover the identifier \texttt{n} for the parameter of a recursive Fibonacci
function. That is either competent reverse engineering or recall of a memorized
idiom cued by a recognizable shape, and no aggregate metric can separate the
two --- because contamination and analysis predict the same aggregate. A model
that has memorized a reference solution can emit correct, readable C from a
maximally mangled \Othree{} input \emph{without performing any repair at all},
and since higher optimization degrades the input more, memorization contributes
proportionally more exactly where the measured effect is largest. Contamination
does not merely inflate such an effect; it can manufacture it.

\paragraph{The control this needs, and what it shows.} We introduce a
within-item ablation we call a \emph{memorization floor}. Refine a function
normally; then refine it again from an input whose identifiers have been
destroyed, and measure how much of the naming gain survives. Because the
comparison is within-item, corpus difficulty cannot contribute to it by
construction. It costs twenty API calls.

Applied to functions written \emph{after} our analysis plan was committed --- so
that no released model could have memorized them --- the floor reports two
things at once, and both matter.

\begin{enumerate}
\item \textbf{Recovery is real.} Refinement recovers identifier signal well
above an arm-matched chance baseline built from its own output vocabulary:
$+0.072$ on novel functions and $+0.110$ on textbook ones for the first
refiner, $+0.105$ and $+0.137$ for a second refiner from a different vendor.
Whatever refinement is doing, it is not producing noise that happens to score.

\item \textbf{But it does not depend on the input we ablate.} Destroying the
input's dataflow changes the naming gain by $+0.001$ (95\% CI
$[-0.026, +0.026]$); removing Ghidra's type prefixes or permuting every
identifier changes it by no more. At the full corpus of each tier, none of the
three input components has a detectable contribution. A second refiner from a
different vendor, given byte-identical inputs under a registration committed
before any call, reproduces that on every contrast: \textbf{twelve contrasts
across two models and two tiers, twelve nulls}.
\end{enumerate}

Readability stays near ceiling under every condition, so a reader is given no
signal that anything has changed. That combination --- real recovery, no
measurable dependence on the ablated input, and undiminished apparent quality
--- is a property this evaluation paradigm has to control for, and the floor is
how we propose controlling for it.

\paragraph{Contributions.}

\begin{enumerate}
\item \textbf{The memorization floor}, a within-item ablation that separates
input-driven recovery from prior-driven recovery, immune to corpus-difficulty
confounds by construction and costing twenty API calls
(\S\ref{sec:floormethod}). We give its manipulation check, without which an
ablation is a claim about information that has not been measured.

\item \textbf{Refinement recovers real signal above chance on uncontaminated
functions}, replicated across two vendors on byte-identical inputs
(\S\ref{sec:recovery}, \S\ref{sec:floor2}). The margin is modest and we report
it as such.

\item \textbf{A bounded null on input dependence}: twelve contrasts, two
models, two tiers, twelve nulls, with the detection bound stated rather than
implied (\S\ref{sec:floor}).

\item \textbf{Arm-matched chance baselines} for generative output. A null built
from reference-vs-reference pairings is the wrong null for any generated arm;
ours sat $0.074$ too high and made real signal read as chance. The correction
costs zero API calls (\S\ref{sec:metric}).

\item \textbf{Execution-specific precision.} A registered re-execution from
byte-identical inputs reproduced every verdict while inflating the null's
detection bound by $44\%$ --- a distinction nobody in this literature reports,
because nobody re-runs (\S\ref{sec:instruments}).
\end{enumerate}

\paragraph{What this paper does not claim.} No registered hypothesis was
confirmed. The readability-gradient hypothesis was refuted; three correctness
hypotheses became untestable because an instrument we registered did not
discriminate on our inputs; and the two primary hypotheses are undecided at the
sample sizes the corpus permits. \S\ref{sec:instruments} reports the five
instrument failures behind those outcomes in full, including two that were
ours rather than the instruments'. We put them in one place, late, because they
are the part most likely to transfer and the part least likely to be read if
scattered.

\section{Related Work}

\paragraph{Decompiler fidelity.} Decompiled C is systematically degraded
relative to source, and not merely cosmetically. \citet{dramko2024taxonomy} give the canonical taxonomy: 15 top-level
categories of fidelity defect identified by open coding across four
decompilers. They also observe that establishing correspondence between source
and decompiled code cannot be done with line numbers or naive heuristics,
because decompilers split expressions, inline statements, and drop code --- an
observation that motivates the alignment-based instrument we adopt. \citet{cao2024evaluating} report that Hex-Rays output recompiles at $30$--$50\%$
while other decompilers manage under $6\%$, with many failures semantically
rooted, and argue on that basis that recompile-and-run methods alone are
unsuitable for assessing semantic accuracy. D-Helix~\citep{zou2024dhelix} shows
Ghidra is sometimes outright wrong, flagging 4{,}515 incorrectly decompiled
functions among roughly 93K and surfacing 17 previously unknown decompiler
bugs. Together these establish both the premise --- Ghidra output leaves real
repair work --- and a constraint on evaluation: Ghidra's own output cannot
serve as ground truth.

\paragraph{Optimization as a fidelity variable.} It is important to be
specific about which axis degrades. Structural fidelity clearly does:
SAILR~\citep{basque2024sailr} traces spurious \texttt{goto} emission to roughly
nine specific compiler transformations, most introduced at \texttt{-O2}, and
finds structural damage at every level, with $17\%$ of spurious gotos present
even at \Ozero{}; \citet{zhou2025rust} corroborates the mechanism
cross-language in Rust. Semantic fidelity does \emph{not} measurably degrade:
\citet{cao2024evaluating} evaluate four decompilers across
\Ozero{}--\Othree{} and \Os{} and find they "all demonstrate certain
robustness across compiler optimization, not showing a significant drop in
effectiveness with the optimization level increasing." Consistent with this,
Agent4Decompile's raw-Ghidra re-executability baseline \emph{rises} from
$18.9\%$ at \Ozero{} to $27.5\%$ at
\Othree{}~\citep{agent4decompile2026}. Every degradation claim in this paper is
therefore scoped to the structural axis. SAILR also cautions against equating
goto-elimination with quality, since original source contains gotos --- a
warning that applies directly to refinement instructed simply to "improve
readability."

\paragraph{LLM refinement.} The pipeline shape we study --- sound lifting by a
traditional decompiler, then LLM repair --- has an established lineage.
DecGPT~\citep{wong2023decgpt} introduced the compiler-in-the-loop pattern over
IDA Pro output, raising recompilation success from $45\%$ to $75\%$.
DeGPT~\citep{hu2024degpt} targets Ghidra with a three-role architecture,
reporting a $24.4\%$ reduction in measured cognitive burden; its MSSC component
symbolically checks that each refinement preserves value behavior and rejects
edits that do not, machinery whose existence is itself evidence that
unconstrained refinement introduces semantic drift, a failure mode also
targeted by feedback-driven multi-turn refinement~\citep{autodecompiler2026}.
LLM4Decompile-Ref~\citep{tan2024llm4decompile} is the best-resourced instance,
and its framing --- Ghidra mangles syntax but preserves underlying logic --- is
the load-bearing assumption of the whole paradigm. D-LiFT~\citep{dlift2025}
contributes the metric structure we adopt: readability credit is awarded only
to output that first passes an accuracy gate.

Recent work sharpens the tension. CoDe-R~\citep{coder2026} identifies degraded
control flow at \Othree{} as the mechanism that limits refinement, while
reporting that its own method still improves over baseline there.
Agent4Decompile~\citep{agent4decompile2026} reports a refinement-baseline table
over identical Ghidra input in which \emph{single-pass} refinement's
re-execution advantage of roughly $+24$ points at \Ozero{} becomes a deficit of
$-5$ at \Othree{}; its own iterative multi-agent method narrows without
reversing; CodeInverter~\citep{codeinverter2025} reports the same
high-at-\Ozero{}, low-at-\Othree{} gradient. Our enhancement arm is single-pass, so the reversing row is the
matching comparator. Meanwhile the readability tables in
Decompile-Bench~\citep{tan2025decompilebench_million} trend the opposite way:
Ghidra's R2I score falls from \Ozero{} to \Othree{} while LLM systems' scores
hold or rise. That table must be cited with a caveat we verified against the
paper: no system it evaluates refines Ghidra or IDA output --- the LLM systems
are end-to-end neural decompilers consuming assembly, so it is
cross-architecture evidence, not same-shape evidence.\footnote{We checked this
directly against the Decompile-Bench paper rather than inheriting it from a
secondary summary, because it is an unusual and load-bearing correction to how
that table is normally read. LLM4Decompile-End and LLM4Decompile-DCBench
consume assembly; Ghidra and IDA appear as standalone traditional-decompiler
baselines, never as an input stage feeding an LLM. The full audit, including
the two claims we could not verify and dropped, is in
\texttt{docs/CITATION\_VERIFICATION.md} in the artifact. The \citet{cao2024evaluating} robustness quotation in this section was verified
the same way, against the paper's own contributions list.}

\paragraph{Symbol recovery and measurement.} Identifier recovery is the
sub-problem our task generalizes, from DIRE~\citep{lacomis2019dire} and
VarBERT~\citep{banerjee2021varbert}, which reports up to $84.15\%$
source-identical variable-name prediction, to
ReSym~\citep{xie2024resym}. Because the readability of decompiled code is to a
large degree the readability of its identifiers, we report naming and
structural gains separately. For readability, R2I~\citep{eom2024r2i} is the only
metric purpose-built for decompiled code; it is relative and cannot score
standalone output, which is why we use a rubric-based judge and report its
limitations rather than claiming a validated absolute scale. For semantic
equivalence we adopt codealign~\citep{dramko2025codealign}, which computes an
equivalence alignment over SSA form and control-dependence information and
yields a graded measure rather than a binary verdict --- pass/fail execution
cannot detect drift on untested paths. \S\ref{sec:codealign} reports that this
instrument did not work on our inputs.

\paragraph{Contamination.} HumanEval-derived decompilation benchmarks are
documented as contaminated~\citep{riddell2024contamination}. To our knowledge no
prior refinement evaluation separates memorization from repair by design. That
gap is what this paper's floor design addresses.

\paragraph{Human factors.} Readability gains must matter to analysts or the
metric measures something without stakes. \citet{yakdan2016johnny} provide the strongest evidence, with user-study
participants solving $3\times$ more tasks under readability-oriented
transformations; observational studies of reverse-engineering
practice~\citep{votipka2020observational} ground rubric design. Our judge is an
LLM, not a human, and we do not claim otherwise.

\section{Method}
\label{sec:method}

\paragraph{Two roles for language models, kept separate.} Language models
appear here in two unrelated capacities. As the \emph{object of study} they are
the refiner and the judge: what they produce is the data, and every detail of
how they were invoked is specified below. As an \emph{authoring aid} they were
used during manuscript preparation and in developing analysis scripts,
including adversarial review passes over near-final drafts; where such a pass
changed a result or its interpretation, the change is recorded with its date
and reasoning in the project log and reported here in the same terms as every
other correction. The author directed the work, verified every reported figure
against the committed artifacts mapped in \texttt{NUMBERS.md}, and is
accountable for the content and the conclusions.

\subsection{Pipeline and corpora}
\label{sec:pipeline}

Each corpus item is one self-contained C program: a target function plus a
\texttt{main()} that \emph{is} its self-test, printing \texttt{PASS} and
exiting~0 when behavior is correct. Items are compiled with gcc 13.3.0 on
x86-64 at \Ozero{}, \texttt{-O1}, \texttt{-O2}, \Othree{} and \Os{}, stripped,
and decompiled with Ghidra 12.1 headless. Each decompiled target function is
then refined by a single stateless API call to Claude Sonnet~5 under a fixed
prompt template, which asks the model to replace generic Ghidra identifiers
with descriptive names, restore types where recoverable, and add explanatory
comments. Both arms --- raw Ghidra and refined --- are evaluated identically.
Sonnet~5 is the refiner throughout except in \S\ref{sec:floor2}, which repeats
the floor design against a second model from a different vendor.

\emph{Naming similarity} compares recovered identifiers against ground truth,
aligned by declaration order, scored by exact-match rate and by cosine
similarity of \texttt{all-MiniLM-L6-v2} embeddings; \S\ref{sec:metric}
establishes what a point of it is worth. \emph{Readability} is scored by a
blind, randomized-order LLM judge (Claude Haiku~4.5, temperature~0) on a
four-part 1--5 rubric. \emph{Functional correctness} is measured by
reassembling each arm's functions into one translation unit, recompiling and
running the self-test.

\paragraph{Two tiers, and why the second exists.} \textbf{T1} is 25
hand-written textbook functions --- the kind of corpus this literature
evaluates on, and the kind a released model has plausibly seen.
\textbf{T3} is 20 functions written \emph{after} the analysis plan was
committed and kept unpublished until this paper, so that no model trained
before their release could have memorized them. Their pre-run existence is
verifiable against a hash manifest committed before the run. T3 is the tier
that carries every claim about uncontaminated recovery; T1 is the comparison.
Tiers are never pooled: no table, figure or abstract number in this paper
averages the two. Because the tiers must be comparable on difficulty for the
cross-tier comparison to mean anything, T3 was built to a difficulty profile
matched against T1 on control-flow depth, loop nesting and operation mix
(Table~\ref{tab:matching}).

\begin{table}[t]
\centering
\small
\begin{tabular}{lccccc}
\toprule
Property & T1 ($n{=}25$) & T3 ($n{=}20$) & diff & overlap & $p$ \\
\midrule
statement count & $6.2 \pm 4.1$ & $5.0 \pm 1.4$ & $-1.2$ & $1.00$ & $0.94$ \\
cyclomatic complexity & $4.6 \pm 3.2$ & $3.5 \pm 1.2$ & $-1.1$ & $0.90$ & $0.46$ \\
max loop-nesting depth & $0.7 \pm 0.5$ & $0.7 \pm 0.5$ & $-0.0$ & $1.00$ & $0.90$ \\
number of parameters & $2.5 \pm 1.8$ & $1.9 \pm 0.6$ & $-0.7$ & $1.00$ & $0.60$ \\
number of local variables & $1.1 \pm 1.0$ & $1.5 \pm 0.8$ & $+0.4$ & $1.00$ & $0.09$ \\
\textbf{distinct identifiers} & $\mathbf{6.6 \pm 3.4}$ & $\mathbf{6.7 \pm 1.7}$ & $\mathbf{+0.1}$ & $\mathbf{1.00}$ & $\mathbf{0.65}$ \\
source LOC & $12.4 \pm 6.8$ & $9.1 \pm 2.9$ & $-3.4$ & $1.00$ & $0.15$ \\
\bottomrule
\end{tabular}
\caption{Difficulty-matching outcome (mean $\pm$ SD). ``Overlap'' is the
fraction of T3 items inside T1's observed range; $p$ is Mann--Whitney,
two-sided. No property separates the tiers.}
\label{tab:matching}
\end{table}

\subsection{The memorization floor}
\label{sec:floormethod}

The floor is the disambiguating design, and it is within-item, so cross-tier
difficulty differences cannot contribute to it by construction. For each item
we re-run refinement on an ablated input in which Ghidra's synthetic
identifiers have been replaced. Two transformations answer to that description
and they are not the same experiment; we state both precisely, because the
distinction is the difference between a result and a null.

\begin{description}
\item[Alpha-renaming.] Build a single bijection over the set of synthetic
identifiers appearing in the function and apply it at \emph{every} occurrence:
draw a permutation $\pi$ of the identifier set once, and rewrite every
occurrence of $v$ as $\pi(v)$. Only the identity permutation is rejected, so
individual names may map to themselves. Every def-use edge survives --- the
result is the same program under a consistent renaming, with Ghidra's
type-encoding prefixes destroyed.

\item[Scrambling.] For each \emph{occurrence} of each synthetic identifier,
draw a replacement uniformly at random from the identifier set,
\emph{independently per occurrence}. The same original name now maps to
different targets at different sites, which is what severs def-use chains.
\end{description}

Three properties of both transforms, stated because
Figure~\ref{fig:scramble} exhibits all of them. The identifier set is
everything matched by the extraction pattern --- variables, parameters,
\emph{and} function and data symbols --- so a function symbol can replace a
variable. Substitution is textual and applies at every occurrence including
declarations and the parameter list. And draws cross prefix classes by design;
the paragraph on type preservation below records the measurement that forced
that choice.

\begin{figure}[t]
\noindent\textbf{Intact} Ghidra output\par\nobreak\vspace{2pt}
\begin{lstlisting}
  char cVar1;
  undefined4 local_10;
  undefined4 local_c;
  local_10 = 0;
  for (local_c = 0; *(char *)(param_1 + local_c) != '\0'; local_c = local_c + 1) {
    cVar1 = *(char *)(param_1 + local_c);
\end{lstlisting}
\vspace{4pt}\noindent\textbf{Alpha-renamed} (what Run~D executed): still a
coherent loop\par\nobreak\vspace{2pt}
\begin{lstlisting}
  char local_c;
  undefined4 local_10;
  undefined4 param_1;
  local_10 = 0;
  for (param_1 = 0; *(char *)(cVar1 + param_1) != '\0'; param_1 = param_1 + 1) {
    local_c = *(char *)(cVar1 + param_1);
\end{lstlisting}
\vspace{4pt}\noindent\textbf{Scrambled} (corrected transform): references
severed\par\nobreak\vspace{2pt}
\begin{lstlisting}
  char param_1;
  undefined4 local_10;
  undefined4 FUN_00101169;
  local_10 = 0;
  for (local_10 = 0; *(char *)(local_10 + local_10) != '\0'; cVar1 = FUN_00101169 + 1) {
    cVar1 = *(char *)(cVar1 + param_1);
\end{lstlisting}
\caption{The two transformations on real Ghidra output
(\texttt{04\_count\_vowels}, \Ozero{}). Under alpha-renaming the loop
counter is consistently \texttt{param\_1} and the loop still reads as a loop:
an analyst --- or a model --- can follow it. Under scrambling the induction
variable, its guard and its increment name three different values, and there
is no chain left to follow. Both draws exhibit the properties stated in
\S\ref{sec:floormethod}: the function symbol \texttt{FUN\_00101169} replaces
a variable (function symbols are in the draw set), \texttt{param\_1}
reappears as a local declaration (substitution covers declarations), and
\texttt{local\_10} is a fixed point of the alpha-renaming permutation (only
the identity permutation is rejected).}
\label{fig:scramble}
\end{figure}

\paragraph{The manipulation check, without which this is not a measurement.}
An ablation is a claim about what information was removed, and a claim about
information is a measurement. Define a \emph{co-reference edge} as an unordered
pair of identifier occurrences naming the same value in the intact input; every
def-use edge is a co-reference edge, so a transform preserving few co-reference
edges cannot preserve many def-use edges. The check is the fraction of those
edges whose two occurrences still carry the same name after the transform.

\begin{table}[t]
\centering
\small
\begin{tabular}{lcc}
\toprule
Transform & def-use survival & chance \\
\midrule
Alpha-renaming (as executed) & $\mathbf{1.000}$ & --- \\
Scrambling, type-preserving & $0.784$ & $0.797$ \\
Scrambling, cross-class (registered) & $\mathbf{0.284}$ & $0.324$ \\
\bottomrule
\end{tabular}
\caption{Manipulation check over the 10 Run~D item-levels. The executed
transform preserved every edge; the corrected one sits at chance.}
\label{tab:manip}
\end{table}

Table~\ref{tab:manip} gives the result: alpha-renaming preserves every edge, as
a bijection must, and the registered cross-class scramble sits at chance. We
recommend this check as mandatory rather than optional. It cost forty lines
over an existing parser, and \S\ref{sec:instruments} reports what happened in
its absence.

\paragraph{Type preservation, decided by measurement.} Ghidra's prefixes encode
type and storage: \texttt{iVar} (signed int), \texttt{uVar} (unsigned),
\texttt{pcVar} (char pointer), \texttt{local\_} (stack), \texttt{param\_}
(argument). Drawing replacements only within a prefix class would preserve that
information and ablate dataflow alone, which is the cleaner experiment. On this
corpus it does not work: these functions carry 2--8 distinct synthetic
identifiers over 2--4 classes, so the modal class is a singleton and admits no
substitution. Type-preserving scrambling leaves $0.784$ of def-use edges intact
and, on 6 of 10 item-levels, leaves \emph{all} of them. We therefore draw
\textbf{across classes}, and state the consequence rather than hide it: the
ablation removes type and storage information along with dataflow.

\paragraph{What the ablation does not remove, and how that bounds the claim.}
Scrambling rewrites identifiers. It leaves the operation vocabulary and the
control structure completely intact --- dereferences, comparison guards,
arithmetic, call structure and literal constants all survive verbatim, and so
does the shape of every loop and branch. Figure~\ref{fig:scramble}'s scrambled
listing is still visibly a loop walking a pointer to a NUL byte. The floor is
therefore a floor against ``the input carries no identifier co-reference and no
type or storage prefixes'', \emph{not} against ``the input carries no
analyzable structure''.

This leaves a competing non-memorization reading we cannot exclude: the model
may be naming from surviving operations alone --- inferring \texttt{count} from
an incremented accumulator --- in which case identifier and type information
were never what it used, and removing them costs nothing without memorization
entering. Distinguishing that needs a fourth condition perturbing operations
while preserving names, which we did not run and register as future work rather
than gesture at. What the present design establishes stands either way: the two
information sources most likely to carry a memorized idiom's surface can both
be destroyed with no measurable effect.

\paragraph{Three conditions, not two.} Because alpha-renaming destroys type
prefixes while preserving dataflow, keeping it as a third condition turns a
two-way comparison into a decomposition:

\begin{center}
\small
\begin{tabular}{lcc}
\toprule
Condition & dataflow & type prefixes \\
\midrule
Intact & preserved & preserved \\
Alpha-renamed & preserved & destroyed \\
Scrambled & destroyed & destroyed \\
\bottomrule
\end{tabular}
\end{center}

Two contrasts follow and answer different questions. \textbf{Intact minus
alpha-renamed} is what Ghidra's type prefixes buy with reference structure held
fixed. \textbf{Alpha-renamed minus scrambled} is the clean estimate of the
dataflow contribution: both conditions carry equally arbitrary names and differ
only in whether the references cohere. Formally, per tier:
\begin{align*}
\Delta_{\text{intact}} - \Delta_{\text{alpha}} &= \text{the type-prefix contribution}\\
\Delta_{\text{alpha}} - \Delta_{\text{scramble}} &= \text{the dataflow contribution}\\
\Delta_{\text{scramble}} &= \text{prior-matching alone.}
\end{align*}
A scrambled-arm gain that stays high means the model's default naming prior
happens to match ground truth. A contribution near zero means the removed
information bought nothing.

\subsection{What a cosine point is worth}
\label{sec:metric}

A number like $+0.060$ is uninterpretable without a scale, and short identifier
embeddings are not well spread. We fix reference points before reporting any
result.

\textbf{Chance --- which must be arm-matched.} The operative null for each arm
is a \emph{permutation null over that arm's own output}: its recovered names for
item $i$ scored against the ground truth of items $j \neq i$, 1{,}000
resamples, seed~0. This matters more than it sounds. Our first chance line
paired each ground-truth identifier with one drawn from a \emph{different}
item's ground truth --- two arbitrary real C identifiers score $\mathbf{0.306}$,
nowhere near $0$ --- but no observed arm pairs ground truth with ground truth,
and model-generated names do not embed like source-idiomatic ones. The
arm-matched lines sit substantially lower (Table~\ref{tab:scale}), and the
difference is not cosmetic: it is what separates ``at chance'' from ``clearly
above chance'' for the refined arm. \S\ref{sec:instruments} reports how we
found this.

\textbf{A useful ceiling.} Each ground-truth identifier is paired with a
hand-written acceptable synonym (\texttt{count}/\texttt{counter},
\texttt{sum}/\texttt{total}), written before inspecting any model output. These
score $\mathbf{0.428}$ on T1. This is the top of the \emph{useful} range, not
of the scale: string identity would score $1.0$, but a reviewer would accept
any of these synonyms. The ceiling shares the vocabulary asymmetry the chance
line had --- it pairs short source-idiomatic ground truth with short
source-idiomatic synonyms, and the refined arm's compound names may be unable
to reach it even when semantically correct --- so the band widths below are, if
anything, too wide, which makes every ``fraction of the band'' claim in this
paper conservative.

\begin{table}[t]
\centering
\small
\begin{tabular}{lcc}
\toprule
Reference point & T1 (textbook) & T3 (novel) \\
\midrule
Raw Ghidra output vs.\ ground truth & $0.239$ & $0.222$ \\
\quad arm-matched chance, Ghidra's names & $0.229$ & $0.212$ \\
Refined output vs.\ ground truth & $0.367$ & $0.309$ \\
\quad \textbf{arm-matched chance, refined names} & $\mathbf{0.257}$ & $\mathbf{0.237}$ \\
Chance, ground truth vs.\ unrelated ground truth & $0.306$ & $0.311$ \\
\textbf{Useful ceiling} (accepted synonyms) & $\mathbf{0.428}$ & $\mathbf{0.435}$ \\
\bottomrule
\end{tabular}
\caption{Reference scale for \texttt{all-MiniLM-L6-v2} identifier
similarity, computed \emph{per tier} because the scale is a property of the
corpus, not of the metric. The useful band for the refined arm --- from
\emph{its own} chance line to the accepted-synonym ceiling --- is $0.171$
wide on T1 and $0.198$ wide on T3, so every delta in this
paper should be read against roughly $0.17$--$0.20$ and not against $1.0$. Each
tier is summarised over its own optimization levels (T1: \Ozero{}--\Os{};
T3: \Ozero{}, \Othree{}). Each arm carries its \emph{own} chance line ---
that arm's actual output vocabulary scored against ground truth from the
wrong item ($1{,}000$ resamples, seed $0$; registered before computation,
producer \texttt{eval/permutation\_null.py}) --- because model-generated
and ground-truth identifiers embed differently, so a single
ground-truth-vs-ground-truth line ($0.306$/$0.311$, retained for
comparison) is the correct null for no observed arm. \textbf{Every arm is
above its own chance line}; the text below takes up how this corrects an
earlier reading. The refined arm's null on T3 has 95\% interval
$[0.223, 0.253]$. The two refined rows are the first refiner's: ``its own
vocabulary'' means the second refiner needs its own line too, and
Table~\ref{tab:null2} gives it. The raw-Ghidra rows are shared by both,
since that arm does not depend on which refiner ran. The synonym ceiling covers
only $47\%$ of ground-truth names (39 had no hand-written synonym) and is
correspondingly less reliable than T1's, which covers all $87$. T3's
refined figure is from the registered re-execution; the raw-Ghidra figure
is deterministic from the committed decompilations.}
\label{tab:scale}
\end{table}

Every delta in this paper should be read against a useful band of roughly
$0.17$--$0.20$, not against $1.0$.

\subsection{Pre-registration}
\label{sec:prereg}

The analysis plan was committed before any API call; git timestamps are the
registration and the full dated log of every deviation, correction and
superseded number ships with the artifact. Three commitments constrain what
follows. Fix rules were written to the log \emph{before} the code implementing
them. Superseded numbers are never overwritten: where an extension changed a
conclusion, both are reported side by side. And instrument parameters are
frozen at their documented defaults, so that no setting can be chosen after
seeing which value flatters a hypothesis.

Two honest qualifications. This is a pre-registration, not a Registered Report:
no reviewer saw the plan before the data existed, so it constrains us but was
not externally validated. And the commit history was rewritten once, after all
analysis was complete, to normalize authorship; content, commit count and every
author timestamp are unchanged, an old-to-new hash mapping ships with the
artifact, and the Zenodo deposit provides third-party timestamping the rewrite
cannot touch.

\subsection{A second refiner}
\label{sec:refiner2}

A floor measured on one model bounds that model. Whether the result is a
property of the paradigm or a quirk of one refiner is not a question one
refiner's data can answer, so we ran the entire floor design again against
\texttt{gemini-3.6-flash}: three conditions $\times$ two levels $\times$ 45
items = 270 calls at temperature~0.

\paragraph{Its registration status, which is weaker than the rest of the
paper's.} The four research questions were fixed before any data existed. This
extension was not: it was registered after the first refiner's results were
known, in the project log, before any code for it was written and before any
call was made. It is a \emph{registered replication}, and we label it that way
throughout rather than let it inherit the original plan's standing. What the
registration fixes is what it always fixes --- the design, the analysis, and a
commitment to report either outcome --- against a result already known.

\paragraph{Byte-identical inputs, which is what makes this matched pairs.} Both
refiners received exactly the same decompiler output. This is checkable rather
than asserted: both runs' raw request snapshots are committed, and all 40 T3
prompts sent to the second refiner are byte-for-byte identical to those sent to
the first. The Ghidra arm is therefore shared between the two analyzes ---
literally the same numbers --- so a cross-refiner difference cannot be a
difference in inputs, and the comparison is paired at the item level rather
than being two independent samples.

\paragraph{Each model at its own floor, not a common setting.} Sonnet~5 rejects
sampling parameters outright, so its variance control is thinking disabled;
Gemini accepts temperature and is set to $0$. For inference-time compute,
Sonnet is disabled and Gemini is at its lowest accepted level, which is not the
same thing. Every call records its thinking-token count and all 270 report
zero, so the claim is the measured one. We compare two refiners each at its own
lowest available setting, and say so rather than describe the runs as
identically configured.

\section{Results}
\label{sec:results}

Runs A (T1, five levels), C (T3), D/D3/D4/D5 (the floor, extended to both full
corpora), a registered T3 re-execution and the 270-call second-refiner
replication completed with zero call failures, at a total measured API spend of
\$6.06 against a \$20 budget --- the corpus, not the budget, is what caps every
sample size here. Every figure maps to a committed producer via
\texttt{NUMBERS.md} in the artifact.

\subsection{Refinement recovers real signal, above chance}
\label{sec:recovery}

The floor is a null, and a null is only informative if the thing being ablated
was measurable to begin with. So we start with what refinement does recover.

Each arm is scored against its \emph{own} arm-matched permutation null
(\S\ref{sec:metric}), recomputed per refiner from that refiner's own output
vocabulary --- reusing the first refiner's would repeat the mis-specification
\S\ref{sec:instruments} reports.

\begin{table}[t]
\centering
\small
\begin{tabular}{llccc}
\toprule
Tier & Arm & observed & arm-matched null & margin \\
\midrule
T1 & raw Ghidra (shared) & $0.239$ & $0.229$ & $+0.010$ \\
T1 & refined, Sonnet~5 & $0.367$ & $0.257$ & $\mathbf{+0.110}$ \\
T1 & refined, Gemini & $0.392$ & $0.255$ & $\mathbf{+0.137}$ \\
\midrule
T3 & raw Ghidra (shared) & $0.222$ & $0.212$ & $+0.010$ \\
T3 & refined, Sonnet~5 & $0.309$ & $0.237$ & $\mathbf{+0.072}$ \\
T3 & refined, Gemini & $0.328$ & $0.223$ & $\mathbf{+0.105}$ \\
\bottomrule
\end{tabular}
\caption{Identifier recovery against each arm's own permutation null
($1{,}000$ resamples, seed~0). The raw-Ghidra rows are shared by
construction --- that arm does not depend on which refiner ran, and both
received identical inputs --- so their exact agreement is a consistency
check on the cross-refiner apparatus rather than a finding.}
\label{tab:null2}
\end{table}

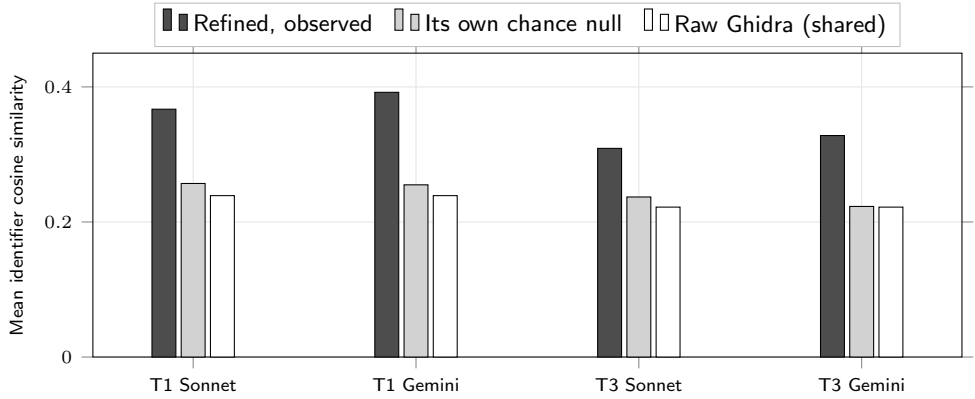
\begin{figure}[t]
\centering
\begin{tikzpicture}
\begin{axis}[
  width=\linewidth, height=5.6cm,
  font=\sffamily\footnotesize,
  ybar, bar width=9pt,
  symbolic x coords={T1 Sonnet, T1 Gemini, T3 Sonnet, T3 Gemini},
  xtick=data, ymin=0, ymax=0.45,
  ylabel={Mean identifier cosine similarity},
  legend style={font=\sffamily\scriptsize, draw=gray!50, fill=white,
                at={(0.5,1.02)}, anchor=south, legend columns=3,
                /tikz/every even column/.append style={column sep=6pt}},
  grid=major, grid style={gray!20}, tick align=outside,
  enlarge x limits=0.15,
]
\addplot[black, fill=black!70] coordinates
  {(T1 Sonnet,0.367) (T1 Gemini,0.392) (T3 Sonnet,0.309) (T3 Gemini,0.328)};
\addlegendentry{Refined, observed}
\addplot[black, fill=gray!35] coordinates
  {(T1 Sonnet,0.257) (T1 Gemini,0.255) (T3 Sonnet,0.237) (T3 Gemini,0.223)};
\addlegendentry{Its own chance null}
\addplot[black, fill=white] coordinates
  {(T1 Sonnet,0.239) (T1 Gemini,0.239) (T3 Sonnet,0.222) (T3 Gemini,0.222)};
\addlegendentry{Raw Ghidra (shared)}
\end{axis}
\end{tikzpicture}
\caption{Refinement recovers identifier signal above chance in both models and
on both tiers. Each refined bar is scored against \emph{its own} arm-matched
permutation null, recomputed from that refiner's output vocabulary
(\S\ref{sec:metric}); a null borrowed from another arm is the wrong comparison
and would move these margins by up to $0.074$. The raw-Ghidra bars are shared
between refiners by construction --- that arm does not depend on which model
ran --- so their agreement is a consistency check rather than a finding. The
margins are $+0.110$ and $+0.137$ on textbook functions and $+0.072$ and
$+0.105$ on functions written after the analysis plan was committed, against a
useful band roughly $0.17$--$0.20$ wide (Table~\ref{tab:scale}). Producer:
\texttt{eval/permutation\_null.py}.}
\label{fig:recovery}
\end{figure}

Figure~\ref{fig:recovery} and Table~\ref{tab:null2} give it. Both refiners clear their own chance lines on both tiers, and by margins that
are modest rather than dramatic: the refined arm sits $+0.072$ to $+0.137$
above chance against a useful band roughly $0.17$--$0.20$ wide, so refinement
buys something in the region of half that band. The raw-Ghidra rows are shared
by construction --- that arm does not depend on which refiner ran, and both
received identical inputs --- so their exact agreement is a consistency check
on the cross-refiner apparatus rather than a finding.

Two things follow. \textbf{On functions written after the analysis plan was
committed}, which no released model could have memorized, refinement still
recovers identifier signal well above chance. And \textbf{the tier ordering
points the same way for both models}: novel functions sit below textbook ones
($+0.072$ against $+0.110$; $+0.105$ against $+0.137$), the direction RQ4
predicted, observed twice on the same items. That is a different measure from
the registered tier comparison in \S\ref{sec:tiercomp}, which tests the naming
\emph{gap} and remains undecided; the agreement is directional, not numerical.

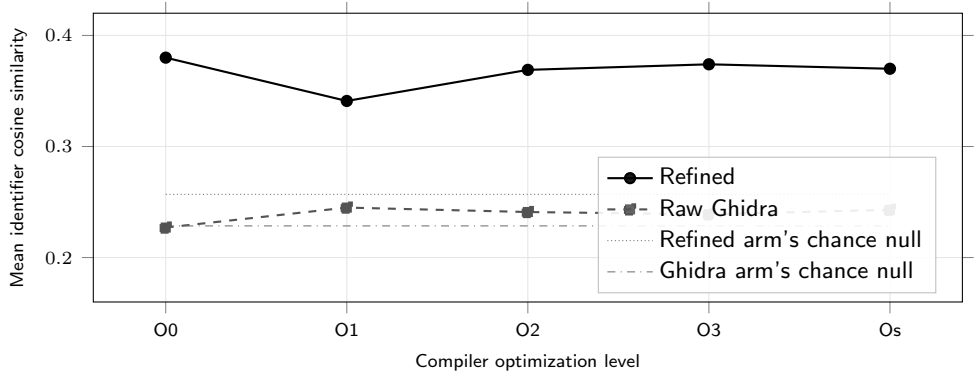
\begin{figure}[t]
\centering
\begin{tikzpicture}
\begin{axis}[emsefig,
  ylabel={Mean identifier cosine similarity},
  ymin=0.16, ymax=0.42,
  legend pos=south east,
]
\addplot[refined] coordinates
  {(O0,0.380) (O1,0.341) (O2,0.369) (O3,0.374) (Os,0.370)};
\addlegendentry{Refined}
\addplot[rawghidra] coordinates
  {(O0,0.227) (O1,0.245) (O2,0.241) (O3,0.239) (Os,0.243)};
\addlegendentry{Raw Ghidra}
\addplot[nullline] coordinates {(O0,0.2569) (Os,0.2569)};
\addlegendentry{Refined arm's chance null}
\addplot[nullline, dashdotted] coordinates {(O0,0.2285) (Os,0.2285)};
\addlegendentry{Ghidra arm's chance null}
\end{axis}
\end{tikzpicture}
\caption{Identifier recovery by optimization level on T1 ($n=25$ per
level), measured as mean cosine similarity of recovered identifiers to
ground truth. Dotted lines are the \emph{arm-matched} permutation nulls of
\S\ref{sec:metric} ($0.2569$ refined, $0.2285$ raw), each built from that
arm's own output vocabulary scored against mismatched references. The
superseded ground-truth-versus-ground-truth chance line of $0.311$ would sit
above the raw arm entirely, which is the error described in
\S\ref{sec:metric}. The gap between arms is flat across levels}
\label{fig:naming}
\end{figure}

Figure~\ref{fig:naming} shows the same picture per optimization level: the
refined arm sits clearly above its own chance line everywhere, the raw arm
barely above its own, and the gap between them does not widen with
optimization.

\subsection{The floor: what that recovery does not depend on}
\label{sec:floor}

Having established that there is signal to ablate, we ablate it.

\begin{table}[t]
\centering
\small
\resizebox{\textwidth}{!}{%
\begin{tabular}{lccc}
\toprule
Contrast & registered $n{=}5$ & pooled $n{=}12$ & \textbf{full corpus} \\
\midrule
\multicolumn{4}{l}{\emph{T1 (textbook), full $n{=}25$}}\\
type prefixes & $+0.060$ $[+0.007, +0.132]$ & $+0.054$ $[+0.007, +0.103]^{*}$ & $\mathbf{-0.005}$ $[-0.052, +0.039]$ \\
dataflow      & $-0.032$ $[-0.074, +0.012]$ & $-0.051$ $[-0.153, +0.069]$ & $-0.003$ $[-0.064, +0.064]$ \\
both          & $+0.028$ $[-0.037, +0.085]$ & $+0.003$ $[-0.082, +0.095]$ & $-0.008$ $[-0.053, +0.043]$ \\
realized MDE, ``both''     & $0.101$ & $0.133$ & $\mathbf{0.071}$ \\
realized MDE, ``dataflow'' & $0.071$ & $0.167$ & $\mathbf{0.094}$ \\
\midrule
\multicolumn{4}{l}{\emph{T3 (novel), full $n{=}20$; re-execution throughout}}\\
type prefixes & $+0.049$ $[+0.014, +0.084]$ & $+0.031$ $[+0.005, +0.056]^{\dagger}$ & $+0.019$ $[-0.004, +0.042]$ \\
dataflow      & $-0.009$ $[-0.048, +0.019]$ & $-0.022$ $[-0.052, +0.007]$ & $-0.020$ $[-0.049, +0.008]$ \\
both          & $+0.040$ $[-0.000, +0.088]$ & $+0.008$ $[-0.031, +0.044]$ & $\mathbf{-0.001}$ $[-0.034, +0.030]$ \\
realized MDE, ``both''     & $0.071$ & $0.056$ & $\mathbf{0.047}$ \\
realized MDE, ``dataflow'' & $0.055$ & $0.044$ & $\mathbf{0.041}$ \\
\bottomrule
\end{tabular}}
\caption{The floor at the full corpus, beside the registered and pooled
analyzes rather than in place of them. Selection was exhaustive --- every
item not already in the $n=12$ subset --- and registered before the run.
Manipulation check on the added item-levels: alpha-rename def-use survival
$1.000$ exactly on all 42, scramble $0.191$ (T1) and $0.187$ (T3) against
chance rates of $0.182$ and $0.153$. \textbf{All six contrasts are null at
the full corpus.} $^{*}$Wilcoxon $p=0.021$ at $n=12$, $p=0.65$ at $n=25$.
$^{\dagger}$Wilcoxon $p=0.034$, the re-execution's nominally significant
T3 prefix contrast; $p=0.123$ at $n=20$. \textbf{The T3 block is the
re-execution at all three $n$}, so its $n=5$ and $n=12$ columns are the
C1R figures of \texttt{docs/RERUN\_COMPARISON.md} rather than the recorded
ones recorded at run time --- the recorded T3 outputs were
destroyed and cannot be extended, so only the re-execution supports a
like-for-like comparison across $n$.}
\label{tab:full}
\end{table}

\textbf{At the full corpus of each tier, none of the three input components has
a detectable contribution on either tier --- six contrasts, six nulls.}
Destroying dataflow and type prefixes together leaves the naming gain
statistically unchanged, and so does each component alone.

The floor null is bounded rather than merely unmeasured. On T3, intact minus
scrambled is $-0.001$ $[-0.034, +0.030]$ with a realised detection bound of
$0.047$, so the full corpus excludes input contributions above roughly $24\%$
of the refined arm's band. We nonetheless \emph{claim} the wider $0.056$ bound
of the $n=12$ re-execution, for a reason \S\ref{sec:instruments} establishes:
the only two executions we have of this measurement differed by $44\%$ in
precision, the full-corpus figure rests on a single execution, and we would
rather quote a bound that survived a replication than a tighter one that has
not faced one.

Figure~\ref{fig:outputs} shows the result at the level of individual
identifiers rather than means, on two rule-selected items.

\begin{figure}[t]
\centering
\small
\begin{tabular}{llll}
\toprule
Ground truth & Intact & Alpha-renamed & Scrambled \\
\midrule
\multicolumn{4}{l}{\emph{T3 \texttt{01\_zigzag\_reset\_accumulate}, \Ozero{} (re-execution)}}\\
\texttt{values} \emph{(par.)} & \texttt{param\_1} & \texttt{arr} & \texttt{arr} \\
\texttt{count}\phantom{s} \emph{(par.)} & \texttt{param\_2} & \texttt{count} & \texttt{count} \\
\texttt{total} & \texttt{running\_sum} & \texttt{running\_total} & \texttt{result} \\
\texttt{index} & \texttt{index} & \texttt{i} & \texttt{i} \\
--- & \texttt{term} & \texttt{term} & \texttt{term} \\
\midrule
\multicolumn{4}{l}{\emph{T1 \texttt{04\_count\_vowels}, \Ozero{} (recorded)}}\\
\texttt{text} \emph{(par.)} & \texttt{str\_ptr} & \texttt{str\_ptr} & \texttt{str} \\
\texttt{count} & \texttt{vowelCount} & \texttt{vowel\_count} & \texttt{vowel\_count} \\
\texttt{i} & \texttt{index} & \texttt{index} & \texttt{index} \\
\texttt{c} & \texttt{currentChar} & \texttt{current\_char} & \texttt{current\_char} \\
\bottomrule
\end{tabular}
\caption{What the model produced for the same function under all three
input conditions, quoted verbatim from \texttt{enhanced.c} in declaration
order. \textbf{Items were selected by rule, not by inspection}: the T3 item
is the first in sorted filename order with all three arms present in the
re-execution trees, and the T1 item is Figure~\ref{fig:scramble}'s
exemplar, so the reader sees the same function whose scrambled input
appears there. The floor result is visible directly. On T3 the ablated arms
name the parameters (\texttt{arr}, \texttt{count}) that the intact arm
leaves as \texttt{param\_1} and \texttt{param\_2}, and the scrambled arm
recovers \texttt{count} and a plausible \texttt{result}/\texttt{i} pair
from an input with every co-reference severed. On T1 all three conditions
produce essentially the same four names. Neither tier shows the degradation
under ablation that a dataflow-driven account predicts. \textbf{The T3 item
reads better than its tier}: two exact matches and three defensible
synonyms in the scrambled column is well above the tier's $3$--$5\%$
exact-match rates. The selection rule was fixed
before inspection and we did not swap the item it picked; the tier means,
not this item, are the evidence. Producer:
\texttt{eval/output\_examples.py}.}
\label{fig:outputs}
\end{figure}

\paragraph{Two effects that did not survive the full corpus.} Both of this
paper's nominally significant contrasts appear at intermediate sample sizes and
disappear at the corpus maximum: T1's type-prefix effect is $+0.054$ with
$p=0.021$ at $n=12$ and $-0.005$ with $p=0.65$ at $n=25$, reversing sign; T3's
prefix contrast is $+0.031$, $p=0.034$ on re-execution at $n=12$ and $+0.019$,
$p=0.123$ at $n=20$. Our registration for the extension committed in advance to
reporting a changed conclusion as prominently as a confirmed one, so we state
it plainly: both are gone, not merely unconfirmed. Neither was a registered
primary contrast, and neither survives a Bonferroni correction across the six.

\paragraph{Power, stated rather than implied.} The registered design's decision
rule for the floor required Wilcoxon $p<0.05$ from a signed-rank test on $n=5$
pairs, whose minimum attainable $p$ is $2/2^{5} = 0.0625$. \textbf{The criterion
was unreachable at the registered sample size regardless of the effect} --- a
defect in our pre-registration, not a property of the data, and the reason the
extensions to $n=12$ and then to the full corpus were registered and run. Table~\ref{tab:floorpower} gives the power analysis. Any
replication needs $n \geq 6$ per tier for $p<0.05$ to be attainable at all. The
corpus caps $n$ at 25 and 20, so the realised bounds in Table~\ref{tab:full}
are close to the best this design can support; a materially tighter bound needs
more items, not more calls.

\begin{table}[t]
\centering
\small
\resizebox{\textwidth}{!}{%
\begin{tabular}{lccc}
\toprule
$n$ per tier & projected MDE (T1) & projected MDE (T3) & min.\ attainable $p$ \\
\midrule
5 \emph{(registered)} & $0.101$ & $0.046$ & $0.0625$ \\
6  & $0.092$ & $0.042$ & $0.0312$ \\
10 & $0.071$ & $0.032$ & $0.0020$ \\
12 & $0.065$ & $0.030$ & $0.0005$ \\
\midrule
25 / 20 \emph{(full corpus, realized)} & $\mathbf{0.071}$ & $\mathbf{0.047}$ & $<10^{-6}$ \\
\bottomrule
\end{tabular}}
\caption{Power for the floor comparison, \emph{projected} from the registered
$n=5$ variance. MDE is the paired minimum detectable effect at 80\% power,
$\alpha=0.05$ two-sided, from the per-item SDs observed at $n=5$ ($0.080$ for
T1, $0.037$ for T3). Minimum attainable $p$ for the signed-rank test is
$2^{1-n}$. \textbf{The realized MDEs at $n=12$ are worse than projected}
($0.133$ for T1, $0.039$ for T3), because the
items added by the extension rule were more variable than the registered
five --- a reminder that power projections from a small pilot are optimiztic
by construction. The final row is realized rather than projected: at the
full corpus the ``both'' MDE is $0.071$ on T1 and $0.047$ on T3
(Table~\ref{tab:full}) --- better than $n=12$ on both tiers, but T1 is
still short of the $0.065$ the $n=5$ variance predicted for $n=12$ alone.
The corpus caps $n$ at 25 and 20, so these are close to the best this
design can do and the tier comparison's power ceiling stands where
\S\ref{sec:threats} puts it.}
\label{tab:floorpower}
\end{table}

\subsection{The floor replicates across vendors}
\label{sec:floor2}

The entire design was executed a second time against
\texttt{gemini-3.6-flash} under the registration and caveats of
\S\ref{sec:refiner2}.

\begin{table}[t]
\centering
\footnotesize
\begin{tabular}{lcc}
\toprule
Contrast (what is removed) & Refiner~1 (Sonnet~5) & Refiner~2 (Gemini) \\
\midrule
\multicolumn{3}{l}{\emph{T1 (textbook), $n=25$}}\\
type prefixes & $-0.005$ $[-0.052, +0.039]$ & $-0.026$ $[-0.085, +0.034]$ \\
dataflow      & $-0.003$ $[-0.064, +0.064]$ & $+0.015$ $[-0.042, +0.079]$ \\
both          & $-0.008$ $[-0.053, +0.043]$ & $-0.011$ $[-0.081, +0.057]$ \\
realized MDE, ``both'' & $0.071$ & $0.098$ \\
\midrule
\multicolumn{3}{l}{\emph{T3 (novel), $n=20$}}\\
type prefixes & $+0.019$ $[-0.004, +0.042]$ & $+0.009$ $[-0.033, +0.062]$ \\
dataflow      & $-0.020$ $[-0.049, +0.008]$ & $-0.002$ $[-0.056, +0.051]$ \\
both          & $-0.001$ $[-0.034, +0.030]$ & $+0.008$ $[-0.030, +0.042]$ \\
realized MDE, ``both'' & $0.047$ & $0.053$ \\
\bottomrule
\end{tabular}
\caption{The floor under two refiners, at the full corpus of each tier.
Percentile bootstrap 95\% CIs, 10{,}000 resamples, seed~0. Refiner~1's
column is Table~\ref{tab:full}'s, unchanged. The two columns reach the same
statistics by different entry points into the same producer --- the
registered full-corpus path for refiner~1, an explicit-arm-files path for
refiner~2, whose trees the former cannot name --- and both delegate to the
same bootstrap, Wilcoxon and MDE code. That the entry points agree is
checked rather than assumed: driven through the explicit-arm path,
refiner~1's recorded arms reproduce its registered T3 column here on all
nine statistics, to every stored digit. Both refiners received
byte-identical inputs (\S\ref{sec:refiner2}), so the Ghidra arm is shared
and the comparison is paired at the item level. \textbf{All twelve
contrasts span zero.} Wilcoxon $p$ for refiner~2:
T1 $0.36$ / $0.84$ / $0.72$, T3 $0.65$ / $0.99$ / $0.43$.}
\label{tab:refiner2}
\end{table}

Table~\ref{tab:refiner2} reports both refiners side by side.

\textbf{Twelve contrasts across two refiners and two tiers, twelve nulls.} No
component of the input we can remove has a detectable contribution for either
model. The result is not a property of one vendor's model.

\textbf{The replication is the weaker of the two measurements, and reading it
the other way round would be the natural error.} Every one of the second
refiner's intervals is wider than the first's and its realised bounds are
larger on both tiers ($0.098$ against $0.071$ on T1, $0.053$ against $0.047$ on
T3). Two models agreeing does not tighten a bound; it adds a second, looser
bound that happens to agree. The bound this paper claims is therefore unchanged
--- the first refiner's $0.056$ --- and the second refiner corroborates the
\emph{verdict}, not the precision.

\paragraph{What we decline to conclude.} The second refiner's margins are
larger and its T1 recovery higher ($0.392$ against $0.367$). We do not read
this as one model being better at reverse engineering. Two refiners at one
prompt each, with no per-model prompt tuning and configurations that are each
model's own floor rather than a matched setting, do not support a ranking. What
the second refiner is evidence for is the thing it was registered to decide,
and on that it is informative precisely because the answer does not depend on
which model is stronger.

\subsection{Tier comparison, readability and correctness}
\label{sec:tiercomp}

\begin{table}[t]
\centering
\small
\begin{tabular}{lcccc}
\toprule
Tier & $n$ & mean per-item gap & sd & median \\
\midrule
T1 & 25 & $+0.1442$ & $0.1495$ & $+0.1595$ \\
T3 & 20 & $+0.0768$ & $0.0514$ & $+0.0646$ \\
\bottomrule
\end{tabular}
\caption{H3-tier, common levels \Ozero{} and \Othree{} only. Tiers never
pooled.}
\label{tab:tier}
\end{table}

Table~\ref{tab:tier} reports the per-item gaps. The registered tier ordering holds --- T3's per-item naming gap is $53\%$ of
T1's, a difference of $+0.0674$ --- but neither deciding test clears $0.05$:
Kruskal--Wallis $H=3.093$, $p=0.0786$; Mann--Whitney $U=327.0$, $p=0.0806$. A
percentile bootstrap CI on the difference is $[+0.0066, +0.1290]$ and excludes
zero. \textbf{We report that tension and do not resolve it in the bootstrap's
favour}: the plan names the two rank tests as deciding, and switching because
the third clears the bar would be choosing the test after seeing the result.
The verdict is undecided. Power caps what a null here could mean anyway --- the
minimum detectable effect is $0.100$, about $69\%$ of T1's observed gap, and
T1's $n=25$ floors it at $58\%$ no matter how many T3 items are written.

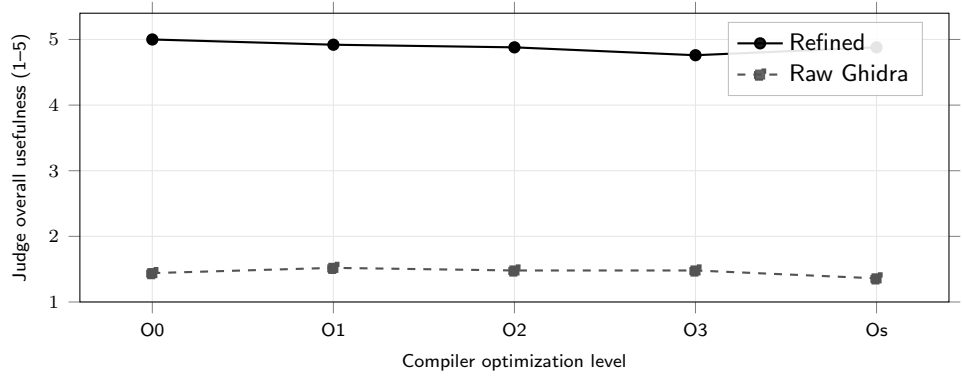
\begin{figure}[t]
\centering
\begin{tikzpicture}
\begin{axis}[emsefig,
  ylabel={Judge overall usefulness (1--5)},
  ymin=1, ymax=5.4,
  ytick={1,2,3,4,5},
  legend pos=north east,
]
\addplot[refined] coordinates
  {(O0,5.00) (O1,4.92) (O2,4.88) (O3,4.76) (Os,4.88)};
\addlegendentry{Refined}
\addplot[rawghidra] coordinates
  {(O0,1.44) (O1,1.52) (O2,1.48) (O3,1.48) (Os,1.36)};
\addlegendentry{Raw Ghidra}
\end{axis}
\end{tikzpicture}
\caption{Blind LLM-judge overall usefulness by optimization level on T1
($n=25$ per level, ungated scores). The gap is large and near-saturated but
does not widen with optimization, which is why H1-readability is refuted
rather than confirmed. \S\ref{sec:judgecal} reports that a
hypothesis-blind human rater agrees on the ordering while placing the gap
$2.8\times$ smaller, so the vertical distance here should be read as
contested in magnitude and not as a validated scale}
\label{fig:readability}
\end{figure}

\paragraph{Readability: the gradient hypothesis is refuted.} H1-readability
predicted the refined-versus-raw readability gap would \emph{widen} with
optimization, on the reasoning that worse input leaves more room to improve. It
does not. Figure~\ref{fig:readability} is two near-horizontal lines separated
by about $3.4$ points at every level. Notably T3's readability gain is
undiminished ($1.23 \to 4.92$) while its naming gain is roughly half T1's ---
consistent with refinement's readability effect being independent of
familiarity while its identifier-recovery effect is not. This is the mechanism
that makes the floor matter: the output looks equally good either way.

\paragraph{Correctness: a null we cannot interpret.} With the reassembly
harness corrected (\S\ref{sec:instruments}), test-pass is $77/125$ for the raw
arm against $75/125$ for the refined arm, with 4 discordant pairs.

\begin{table}[t]
\centering
\small
\begin{tabular}{lcccc}
\toprule
& \multicolumn{2}{c}{recompiles} & \multicolumn{2}{c}{test passes} \\
\cmidrule(lr){2-3}\cmidrule(lr){4-5}
Level & Ghidra & LLM & Ghidra & LLM \\
\midrule
\Ozero{}    & 25 & 24 & 17 & 17 \\
\texttt{-O1} & 22 & 22 & 16 & 16 \\
\texttt{-O2} & 20 & 20 & 17 & 16 \\
\Othree{}   & 19 & 18 & 17 & 16 \\
\Os{}       & 14 & 14 & 10 & 10 \\
\midrule
Total       & 100 & 98 & \textbf{77} & \textbf{75} \\
\bottomrule
\end{tabular}
\caption{Functional correctness, $n=25$ per level. 4 discordant pairs of 125.}
\label{tab:corr}
\end{table}

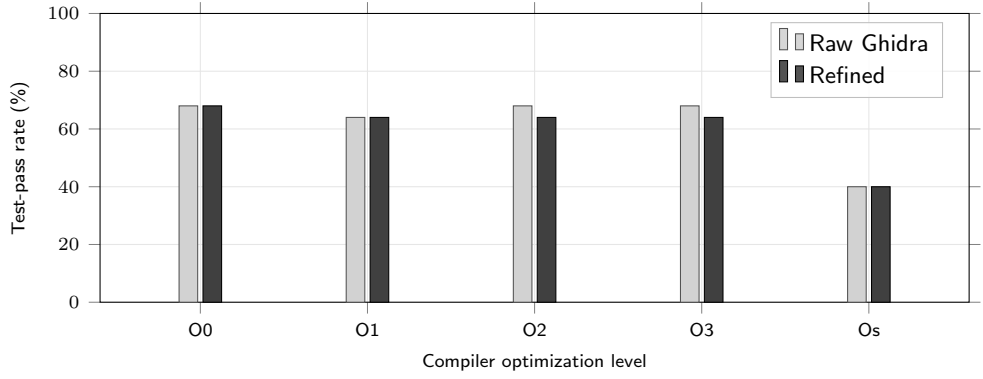
\begin{figure}[t]
\centering
\begin{tikzpicture}
\begin{axis}[emsefig,
  ybar,
  bar width=7pt,
  ylabel={Test-pass rate (\%)},
  ymin=0, ymax=100,
  enlarge x limits=0.15,
  legend pos=north east,
]
\addplot[fill=gray!35, draw=gray!65!black] coordinates
  {(O0,68.0) (O1,64.0) (O2,68.0) (O3,68.0) (Os,40.0)};
\addlegendentry{Raw Ghidra}
\addplot[fill=black!75, draw=black] coordinates
  {(O0,68.0) (O1,64.0) (O2,64.0) (O3,64.0) (Os,40.0)};
\addlegendentry{Refined}
\end{axis}
\end{tikzpicture}
\caption{Functional correctness by optimization level on T1, as the
percentage of the $25$ items per level whose reassembled output recompiles
and passes its own self-test. Unconditional rates: an item that fails to
recompile counts as a failure rather than being excluded. The arms differ by
a single item at \texttt{-O2} and at \Othree{}, and not at all at the other
three levels, totalling $77$ versus $75$ of $125$ with four discordant
pairs. These are the corrected figures;
\S\ref{sec:correction} reports the harness bugs that had made this
comparison read as a $20$-point deficit against refinement}
\label{fig:testpass}
\end{figure}

Table~\ref{tab:corr} gives the breakdown by level and Figure~\ref{fig:testpass} the same data by arm.
Four discordant pairs in 125 is not demonstrated equivalence; it is a
near-degenerate axis with almost no power to detect a real effect of moderate
size. We report it descriptively and draw no conclusion from it.

\section{Instrument Validity}
\label{sec:instruments}

This study set out to measure refinement and spent most of its evidence
measuring its own instruments. Five measurements we registered turned out to
be measuring something other than what we intended. We report them together,
here, rather than scattered through the results, because they are the part
most likely to transfer to other pipelines and the part least likely to be
read if distributed. Two of the five are failures of instruments; two are
failures of ours; one is a property of the pipeline nobody in this literature
measures.

\begin{table}[t]
\centering
\small
\begin{tabular}{lllp{5.6cm}}
\toprule
\textbf{RQ} & \textbf{Hypothesis} & \textbf{Verdict} & \textbf{Basis} \\
\midrule
RQ4 & H3-tier \emph{(primary)} & Undecided & Ordering holds ($+0.144$ vs $+0.077$) but Kruskal--Wallis $p=0.079$, Mann--Whitney $p=0.081$; CI spans the MDE \\
RQ3 & H3-floor \emph{(primary)} & Undecided & Registered criterion unattainable at $n=5$ \\
RQ1 & H1-readability & \textbf{Refuted} & Gap is flat across levels, not widening \\
RQ2 & H1-correctness & Untestable & codealign $I=0.092 < 0.20$ \\
RQ2 & H1-divergence & Untestable & Requires H1-correctness \\
RQ2 & H2 & Untestable & Same instrument \\
\bottomrule
\end{tabular}
\caption{All registered hypotheses, in the plan's order, with the research
question each operationalizes. Run~B (T2) was deferred,
so the monotone three-tier ordering was never evaluated. RQ5 appears in no
row: it was not registered, and no hypothesis was written for it.}
\label{tab:hyp}
\end{table}

Table~\ref{tab:hyp} states the consequence in the plan's own terms.
\textbf{No registered hypothesis was confirmed.}

\subsection{Harness bias is directional, and it favours the raw arm}
\label{sec:correction}

Our first correctness measurement showed the refined arm losing to raw Ghidra
output by roughly twenty points. It was an artifact of our own reassembly
harness, and the failure modes were not random: \textbf{every one of them
penalised the model for following the registered prompt.}

\begin{table}[t]
\centering
\small
\begin{tabular}{p{5.2cm}p{4.2cm}c}
\toprule
Bug & Penalized the model for & Pairs \\
\midrule
Reassembler could not link a renamed function & obeying prompt instruction 1 (rename identifiers) & 25 \\
\texttt{extract\_prototype} flattened the leading block comment, leaving it unterminated and swallowing following definitions & obeying prompt instruction 3 (add comments) & 13 \\
Fence stripping left stray \verb|```| lines & emitting markdown & 4 \\
\bottomrule
\end{tabular}
\caption{Harness fragility is directional: it runs systematically against the
behaviors refinement is asked to produce. Each bug damaged only the refined
arm, and each did so \emph{because} the model complied with a prompt
instruction. None touched the control arm.}
\label{tab:bugs}
\end{table}

Table~\ref{tab:bugs} lists them.
The prompt asks for descriptive names, so the reassembler could not link a
renamed function; it asks for explanatory comments, so a prototype extractor
that flattened block comments truncated declarations; the model emitted
markdown fences, which stripping missed. Each bug fires only when refinement
does its job. Two further bugs surfaced later, one arm-neutral and therefore
invisible to every between-arm check we had registered.

The transferable point is the shape rather than the count. A reassembly harness
sits downstream of exactly the behaviors refinement is asked to produce, so
its failures are systematically anti-correlated with treatment success. A
between-arm difference is the natural check and it cannot catch this; what
caught ours was a manipulation check on \emph{absolute} plausibility --- asking
whether a 20-point deficit was credible at all, rather than whether it was
larger than zero. We recommend reassembly harnesses be treated as a primary
suspect for correctness deficits in refinement studies, and that absolute
plausibility be checked before any between-arm comparison is believed.

\subsection{An LLM judge tracks a human's ranking but not the size of the gap}
\label{sec:judgecal}

\begin{table}[t]
\centering
\small
\begin{tabular}{lccccc}
\toprule
& \multicolumn{2}{c}{mean rating} & \multicolumn{2}{c}{} & \\
\cmidrule(lr){2-3}
Rubric dimension & raw & refined & gap & Spearman & mean $|$diff$|$ \\
\midrule
\multicolumn{6}{l}{\emph{Human rater (blind, $n=20$ pairs)}}\\
Naming clarity      & $2.05$ & $4.60$ & $+2.55$ & $0.84$ & $0.65$ \\
Structural clarity  & $3.65$ & $4.25$ & $+0.60$ & $0.37$ & $0.88$ \\
Comment usefulness  & $1.45$ & $4.90$ & $+3.45$ & $0.91$ & $0.30$ \\
Overall usefulness  & $3.00$ & $4.25$ & $\mathbf{+1.25}$ & $0.77$ & $1.20$ \\
\midrule
\multicolumn{6}{l}{\emph{LLM judge, same 20 pairs}}\\
Naming clarity      & $1.10$ & $4.65$ & $+3.55$ & & \\
Structural clarity  & $3.15$ & $4.50$ & $+1.35$ & & \\
Comment usefulness  & $1.00$ & $4.85$ & $+3.85$ & & \\
Overall usefulness  & $1.30$ & $4.75$ & $\mathbf{+3.45}$ & & \\
\bottomrule
\end{tabular}
\caption{Human calibration of the LLM judge. Spearman and mean absolute
difference are human-vs-judge, over 40 ratings per dimension. Test--retest
reliability from the four repeated pairs: Spearman $0.80$, mean $|$diff$|$
$0.94$, exact agreement $25\%$.}
\label{tab:judgecal}
\end{table}

Table~\ref{tab:judgecal} gives the comparison. Against one human rater blind to our hypotheses, the judge ranks arms as the
human does ($\rho=0.77$; both prefer refinement in $18$--$20$ of $20$ pairs)
but reports a gap $2.8\times$ larger, and the disagreement sits almost entirely
in how the two score \emph{raw decompiler output} rather than refined output.
Neither instrument is validated, so this is a measured divergence rather than a
demonstrated judge error --- but at least one of the two is wrong about that
baseline, and rank correlation alone would have hidden it either way. If the
human is closer to right, every comparison drawn against a decompiler baseline
in this literature is inflated. \textbf{LLM judges should be calibrated on
effect size, not only on ranking.}

The calibration is one rater on 20 pairs and unregistered; per-item
disagreement is close to that rater's own test--retest noise, so we rest
nothing on individual items and report only the systematic gap difference,
consistent across $17/20$ pairs and all four rubric dimensions.

\subsection{We registered a correctness instrument without checking it worked
on our inputs}
\label{sec:codealign}

\begin{table}[t]
\centering
\small
\begin{tabular}{lcc}
\toprule
Metric & Value & Registered threshold \\
\midrule
Applicability (both arms scorable) & $109/125 = 87.2\%$ & $\geq 80\%$: passes \\
Informative-pair rate $I$ & $\mathbf{0.092}$ & $\geq 0.20$: \textbf{fails} \\
Ceiling rate $C$ & $0.028$ & $< 0.60$ \\
\bottomrule
\end{tabular}
\caption{codealign gate. Applicability passes; discrimination does not.}
\label{tab:codealign}
\end{table}

Table~\ref{tab:codealign} gives the registered gate.
codealign~\citep{dramko2025codealign} is a graded instruction-level equivalence
checker, and we registered it as our primary correctness measure. Its
informative-pair rate on our data is $0.092$ against a threshold we registered
at $0.20$ before seeing any data, and mean coverage falls from $0.31$ at
\Ozero{} to $0.007$ at \Othree{}. Three of our six hypotheses became untestable
as a pre-registered consequence.

\textbf{Every failure we hit lies inside limitations codealign's own paper
documents}, so this is a finding about our procedure rather than about the
tool. Its paper excludes from its own evaluation ``functions which contain
features codealign does not currently support, such as \texttt{\#ifndef} macros
and \texttt{goto} statements'', and states that ``instructions must have at most
one control dependency''. Ghidra emits \texttt{goto} and labels by design when
structured control flow cannot be recovered. The authors exclude such
functions; we did not.

The collapse itself is a control-dependence cascade rather than a gradual loss
of alignable structure: loop rotation at \texttt{-O1} and above introduces a
guarding \texttt{if} in the candidate with no counterpart in the reference, it
fails to align, and every instruction control-dependent on it fails with it.
One item shows it in isolation --- a five-line \texttt{while} loop computing a
GCD aligns $6/6$ reference instructions at \Ozero{} and $0/6$ at \Othree{},
both decompilations correct and differing only in loop shape.\footnote{Mechanism
established in correspondence with codealign's authors, whom we thank. A
standalone reproduction is available with the artifact.}

Because the mechanism implicates a registered parameter, we checked whether the
verdict depends on it: re-running all 250 pairs with control dependence
disabled raises the informative-pair rate from $0.092$ to $0.193$ --- still
below the $0.20$ registered before any data existed, by one pair of $109$. The
verdict is unchanged. We report this as an exploratory sensitivity analysis and
not as a substitute for the registered figure.

\textbf{The generalization is cheap to state and would have been cheap to
act on.} Applicability is a property of an instrument \emph{on an input class},
and a registration that does not establish it is registering a hope. One
\Othree{} item, run before registration, would have exposed all of this in an
afternoon --- at which point the design could have excluded
\texttt{goto}-bearing functions as codealign's own evaluation does, chosen a
different instrument, or registered the pass/fail axis as primary instead.

\subsection{Two errors of ours, and how each was caught}
\label{sec:ourerrors}

\paragraph{A chance baseline built from the wrong vocabulary.} Our first null
paired ground-truth identifiers with ground-truth identifiers from other items.
No observed arm does that, and model-generated names embed with lower mean
similarity to everything, so the correct arm-matched null sits $0.074$ lower.
A released draft read genuinely above-chance recovery as chance on that basis.
The correction costs zero API calls and changes a headline in either direction,
which is why we argue it is the only defensible chance line for generative
output.

\paragraph{An ablation that did not ablate.} Run~D as first implemented was a
consistent bijection, which preserves every def-use edge; measured def-use
survival was $1.000$. A full draft of this paper reported a dataflow result on
that basis. The manipulation check of \S\ref{sec:floormethod}, which the design
had omitted, is what caught it. The correction cuts both ways: it rescued the
novel-tier claim, now supported by an ablation that demonstrably removes what
it says it removes, and it cost the textbook-tier claim, which did not survive
the same treatment. \textbf{Every ablation should ship a manipulation check
quantifying what it removed}, reported next to the effect it licenses.

\paragraph{The failure mode neither pre-registration nor re-execution can
catch.} Both of these were caught by outside reads of near-final drafts, not by
anything the registered study contained. That is not coincidence but a
structural property. Pre-registration and re-execution are both \emph{faithful}
mechanisms: registration fixes the question before the data can bias the
answer, and re-execution confirms the pipeline computes what it says. A
faithful mechanism cannot detect that the question itself is wrong, because it
reproduces the wrong comparison exactly, now with authority. Every number in
this paper was correctly computed, and two of them were correct answers to the
wrong question. Against that failure mode we know of one defence --- a reader
who has not spent months inside the design's assumptions --- and we would now
budget for such a read as deliberately as we budget for a re-execution.

\subsection{Reported precision belongs to one execution}
\label{sec:reexec}

After the T3-side per-item artifacts were destroyed by a file-handling accident
\emph{after} analysis was complete, we re-executed the entire T3 side --- 128
calls --- from inputs verified byte-identical, under a registration committed
before any call, with recorded and re-run numbers reported side by side
regardless of agreement.

The outcome cuts both ways. Every registered hypothesis verdict reproduces; the
central null reproduces in sign and nullity at all three sample sizes; and the
T1 control recomputes to four decimals, confirming an identical measurement
pipeline. But the \emph{precision} does not reproduce: the pooled detection
bound inflates from $0.039$ to $0.056$, a $44\%$ move, and a secondary contrast
crosses nominal significance in one execution and not the other. At the
registered $n=5$ the same contrast moves by $0.049$ between executions ---
comparable to the $0.056$ bound this paper claims --- shrinking with $n$ and
washing out at the full corpus.

\textbf{An LLM-in-the-loop pipeline re-executed from identical inputs thus
reproduced every pre-registered verdict while moving estimates enough to break
a claimed bound.} That is simultaneously a validation of verdict-level
pre-registration and a caution against quoting tight bounds on nulls, or
unregistered significance, from a single execution. Ours cost \$1.19. Nobody in
this literature re-runs; this is what one re-run showed.

\section{Threats to Validity}
\label{sec:threats}

\paragraph{Scope.} Single compiler (gcc 13.3.0), single decompiler (Ghidra
12.1), single architecture (x86-64), one prompt per model, textbook-scale
single functions with self-tests. Results may not transfer to real-world
binaries, other decompilers, or larger functions. \Os{} is reported
descriptively and excluded from every trend test. The floor is the one finding
measured on two refinement models; everything else rests on the first refiner
alone.

\paragraph{The corpus is small, and the corpus is the binding constraint.} The
floor's detection bounds are $0.047$ (T3) and $0.071$ (T1) at the corpus
maximum, roughly a quarter of the refined arm's useful band. The corpus caps
$n$ at 20 and 25, so a materially tighter bound needs more items rather than
more calls --- the full-corpus extension cost \$1.04. Budget was never the
binding constraint and we would rather say so than let a reader supply a worse
explanation.

\paragraph{The ablation conflates its target with idiom recognition.}
Scrambling destroys dataflow, prefix types and the surface cues that would let
a model recognize a memorized idiom, all at once, so a contrast is not
attributable to any one of them alone. This bounds the T1 comparison. The T3
null does not depend on it, since there nothing is detectable under any
condition. The competing ``naming from operations alone'' reading
(\S\ref{sec:floormethod}) remains open and needs a fourth condition we did not
run.

\paragraph{Two models is a small sample of the population that matters.} The
replication distinguishes ``one model's quirk'' from ``not one model's quirk''
and no more. Both models ran at one prompt each and at each model's own lowest
inference-time setting rather than a common one, so neither prompt sensitivity
nor reasoning-budget sensitivity is bounded by anything here. The extension was
also registered after the first refiner's results were known, which is why we
call it a registered replication rather than folding it into the
pre-registration.

\paragraph{We wrote the T3 items ourselves, knowing the hypothesis.}
\S\ref{sec:pipeline} addresses the \emph{difficulty} confound; it does not
address demand characteristics. An author who expects to show that refinement
cannot name unfamiliar functions may, without intending to, write functions
whose names are hard to guess from behavior. Nothing in our design excludes
this. The cheap empirical bound --- have readers blind to the source propose
names from behavior alone, and compare the rate across tiers --- is not run
here; the stronger fix is to have probe items written by someone who does not
know the hypothesis, and we recommend it for any replication.

\paragraph{The judge is an LLM and shares a developer with the model under
evaluation.} Blinding and order randomization address presentation bias, not
that. \S\ref{sec:judgecal} reports a measured divergence against one
unregistered volunteer rater without reverse-engineering experience; an
experienced panel is the proper version and was not run. Readability magnitudes
should be read as contested rather than as bounded in a known direction.

\paragraph{Run~B was deferred.} No T2 corpus was built, so the registered
monotone three-tier ordering was never evaluated. $\Delta(T1) > \Delta(T3)$
must not be read as if a monotone ordering had been tested.

\section{Conclusion}
\label{sec:conclusion}

We set out to measure how the readability/correctness trade-off in LLM
refinement of decompiler output scales with compiler optimization. We did not
succeed: the graded semantic instrument we registered does not discriminate on
our inputs, which made three of six hypotheses untestable, and the readability
hypothesis was refuted because the gap is flat and near-saturated rather than
widening.

What the study does establish is narrower and, we think, more useful. On
functions written after the analysis plan was committed, refinement recovers
identifier signal well above an arm-matched chance line --- and that recovery
does not measurably depend on the input we can ablate. Destroying dataflow
changes the naming gain by $+0.001$ (95\% CI $[-0.026, +0.026]$); removing type
prefixes or permuting names changes it by no more; and a second refiner from a
different vendor, on byte-identical inputs, reproduces the null on all six of
its contrasts. Readability stays near ceiling throughout, so a reader is given
no way to tell. Since refined output is what an analyst actually reads, and
since it looks equally confident either way, this is a property the evaluation
paradigm has to control for.

We are deliberate about what this does \emph{not} show. It is a bounded null,
not a demonstration that nothing is happening: contributions smaller than
$0.056$ are invisible to us, and the ablation leaves the operation vocabulary
intact, so naming from operations alone remains a live competing reading. The
replication does not tighten anything --- its intervals are wider throughout,
so the bound we claim remains the first refiner's --- and one prompt per model
leaves prompt sensitivity untested.

\paragraph{Five recommendations.} A \textbf{memorization floor} should be a
standard control in refinement evaluation: it is within-item and therefore
immune to corpus-difficulty confounds, it costs twenty API calls, no aggregate
metric substitutes for it, and where a floor result is the headline it should
be run against a second model from a different vendor on byte-identical inputs,
which cost us \$0.74. \textbf{Every ablation should ship a manipulation check}
quantifying what it removed, reported beside the effect it licenses; ours cost
forty lines and caught a defect that had survived an entire study and a full
draft. \textbf{Chance baselines for generative output must be arm-matched},
built from the system's own output vocabulary rather than from
reference-versus-reference pairings; ours differed by $0.074$ and the error
survived every earlier review. \textbf{Graded instruments should carry an
applicability report on the specific inputs used}, registered in advance --- a
null result is otherwise indistinguishable from an instrument that never had
signal on those inputs, and one item run before registration would have told us.
And \textbf{LLM-in-the-loop evaluations should re-execute at least once} under
a registration committed before the re-run: ours cost \$1.19, reproduced every
verdict, and showed that the precision we had been quoting belonged to one
execution rather than to the pipeline.

\paragraph{Artifact.}
\label{sec:artifact}
The corpus (both tiers), pipeline, pre-registered analysis plan, the complete
dated research log including every correction and deviation, raw API responses
for the T1 runs, the registered T3 re-execution and all 270 calls of the
second-refiner replication, and a per-number provenance map are archived at
\url{https://doi.org/10.5281/zenodo.21968878}. The original T3-side raw
responses and per-item result files were destroyed by a file-handling accident
after analysis; the log records the loss and exactly what survives, and the
re-execution's side-by-side comparison against every recorded T3 figure ships
with it. The T3 probe items are published here, and their pre-run existence is
verifiable against a hash manifest committed before the run. Publishing them is
one-way: they cannot serve as an uncontaminated probe for any later-trained
model, and a replication wanting a clean probe needs new items and a new
commitment.

\backmatter

%
\section*{Declarations}

\paragraph{Funding.}
No funding was received for conducting this study. The author received no
grant, studentship, internship, institutional support, or vendor credits of
any kind. The study's entire computational cost was \$6.06 in API charges,
paid by the author personally. The affiliation above identifies where the
author is enrolled as a student; the institution had no role in the design,
execution, analysis, or reporting of this work, which the author conducted
independently.

\paragraph{Competing interests.}
The author declares no competing interests, financial or non-financial. The
study evaluates commercial models produced by Anthropic and by Google; the
author has no employment, consulting, financial or other relationship with
either company, received no support, discounts or credits from either, and
paid list price for all API usage reported here. The second refiner
(\S\ref{sec:refiner2}) was registered to run on Google's free tier and moved
to the paid tier by a filed amendment when the free tier's daily quota proved
incompatible with the registered design; that run was billed at list price
like every other.

\paragraph{Ethics approval and consent to participate.}
The judge-calibration analysis (\S\ref{sec:judgecal}) involved one adult
human participant rating twenty pairs of source code. \textbf{No
institutional ethics review was sought for this study.} The author states
the protocol so that a reader may judge that decision: the study was
non-interventional; it collected no personal, sensitive or demographic
data; it involved a single adult volunteer personally known to the author,
who gave informed consent to participate and to publication and was free to
withdraw at any point; the released dataset contains rating values keyed to
pair identifiers and nothing that could identify him; and the only
information withheld was the study's hypotheses, which the blinding
required.

\paragraph{Consent for publication.}
The participant consented to publication of his aggregated and per-pair
ratings in anonymous form.

\paragraph{Data and code availability.}
The corpus (both tiers), the full pipeline, the pre-registered analysis
plan, the complete dated research log, per-run configuration and raw API
responses for every run that survives, and a per-number provenance map are
available in the replication package archived at
\url{https://doi.org/10.5281/zenodo.21968878} (all versions; v1.1.0 current at
the time of writing),
released under CC~BY~4.0 for data and prose and the MIT licence for code.
Three limits on that package are stated so a reader knows what is
\emph{not} in it. The manuscript source is excluded and its rights
reserved, the paper being under submission when the deposit was made; the
provenance map is included, since this paper cites it. The original T3-side
raw responses and per-item result files were destroyed by a file-handling
accident after analysis and before write-up (\S\ref{sec:artifact}), so T3
figures rest on the registered re-execution rather than on the original
run. And publishing the T3 probe corpus is irreversible: those items cannot
serve as an uncontaminated probe for any model trained after their release.

\paragraph{Author contributions.}
The single author designed the study, wrote the pre-registration, built the
pipeline and evaluation code, executed all runs, performed the analysis, and
wrote the manuscript.

\paragraph{Use of generative AI in the preparation of this manuscript.}
Generative AI was used in two distinct roles, which we separate to avoid
conflating them. First, as the \emph{object of study}: the refinement and
judging models are the subject of the experiments and are documented in
\S\ref{sec:method}. Second, as an \emph{authoring aid}: large language
models were used during manuscript preparation and analysis-script
development, including review passes over near-final drafts. The author
directed all of this work, verified every number
against the committed artifacts listed in \texttt{NUMBERS.md}, and takes
full responsibility for the content, the analysis and the conclusions. No
AI system is or could be an author of this paper.

\bibliography{refs}

\end{document}